\documentclass[a4paper,11pt]{article}
\usepackage{pos}

\begin{document}

\newcommand{\cM}{\mathcal{M}}
\newcommand{\cR}{\mathcal{R}}
\newcommand{\cN}{\mathcal{N}}
\newcommand{\cD}{\mathcal{D}}
\newcommand{\cF}{\cal{F}}
\def\beq{\begin{equation}}
\def\eeq{\end{equation}}
\def\beqa{\begin{eqnarray}}
\def\eeqa{\end{eqnarray}}

\title{Some Properties and Uses of the Species Scale}
%% \ShortTitle{Short Title for header}

\author*[a,b]{Luis E. Ib\'a\~nez}
%\author[a,b]{Second Author}

\affiliation[a]{Instituto de F\'isica Te\'orica UAM-CSIC ,\\
  c/ Nicolas Cabrera 11-13, Cantoblanco, 28034 Madrid}

\affiliation[b]{Departamento de F\'isica Te\'orica, \\
Universidad Aut\'onoma de Madrid, Cantoblanco, 28034 Madrid}

%\emailAdd{f.author@inst.edu}
%\emailAdd{s.author@univ.country}

\abstract{  The 'Species Scale' has proved to be an important concept when studying consistent effective actions 
in Quantum Gravity.
This is a short summary of my contribution to the Corfu Summer Institute in September 2025,  in which I covered  two topics,  both related 
in different ways  to the fact that the Species Scale is moduli dependent. In the first, based on work done 
in collaboration with C. Aoufia and A. Castellano, we  show how the  one-loop Wilson coefficients  ${\cF}_n^{(d)}$ multiplyiing BPS protected ${\cal R}^{2n}$ 
operators obey Laplace-like eigenvalue differential equations of the form
$\cD^2_{\bf {\cal M}} {\cF}^{(d)}_n   = \eta_d\,  {\cF}^{(d)}_n$. This is true both for $n=2$ with 32 and 16 SUSY generators in 
10,9,8 dimensions  and theories with 8 SUSY generators in 6,5,4 dimensions $(n=1)$. We argue that this fact is at the root of
some Swampland conjectures put forward in the past, like bounds on  the dumping rates for the tower scales  and the
exponential behaviour in the  Swampland Distance Conjecture. For the second topic, based on work done in
collaboration with G.F. Casas, we discuss the one loop potential of the no-scale moduli in GKP-like Type IIB 4d orientifolds.
To compute this potential we sum both over light and heavy (tower) modes using the Species Scale as a UV cut-off. 
We find  a generic form $V_{1-loop}\sim g^2m_{3/2}^2M_p^2(g^{i{\bar i}}(\partial_i\Lambda)(\partial_{\bar i}\Lambda))/\Lambda^2$,
with $\Lambda$ the Species Scale.
This has minima at the {\it desert points} in moduli space and exponentially decreases at large moduli, with a dS hill in between. 
We argue that this potential may lead to the stabilisation of some or all Kahler moduli at the desert points in 4d Type IIB orientifolds
of phenomenological interest.}

\FullConference{Proceedings of the Corfu Summer Institute 2025 "School and Workshops on Elementary Particle Physics and Gravity" (CORFU2025)\\
27 April - 28 September, 2025\\
Corfu, Greece\\}

%% \tableofcontents

%\begin{document}
\maketitle

\section{ The Species Scale}

It seems natural to consider the Planck scale $M_{pd}$ as the scale at which Quantum Gravity (QG) 
becomes important in a theory of $d$-dimensions. However since first pointed out  by Gia Dvali \cite{Dvali:2007hz,Dvali:2007wp,Dvali:2008ec} and
collaborators, it has become clear that the notion of QG fundamental scale is slightly more subtle, particularly in the
presence of a large number of degrees of freedom $N$, like those of a KK  or string tower, which appear when the moduli  go to points at infinite distance.
In those limits the {\it Swampland Distance Conjecture} (SDC) \cite{Ooguri:2006in}
states that towers of such particles must exist, 
with lightest state mass in the tower exponentially dumping, $m_t\sim m_{0t}exp(-\lambda_t\Delta\phi)$,with $\Delta \phi$ some
geodesic distance in moduli space. The spectra  of towers in string theory takes typically the following general form for the mass of the $n$-th level state
(see e.g.\cite{Castellano:2021mmx,Castellano:2022bvr})
\beq
m_n^2 \ =\ n^{2/p} m_0^2 \ ,\ n=0,...,N       \ .
\label {torre}
\eeq
Thus, e.g., for a single KK tower one has $p=1$ and the familiar relation $m_n^2=n^2m_0^2$ follows. The case $p=2$ would correspond e.g., to an effective tower 
for two KK towers.  Concerning string towers,  although one would also naively say that one has  $p=2$, it turns out that a more appropriate
description is $p=\infty$ 
since the degeneracy at each level grows like $e^{\sqrt{n}}$
\cite{Castellano:2021mmx,Castellano:2022bvr}.  
%In particular note that for $p\rightarrow \infty$ 
%eq.(\ref{torrespecies}) yields $\Lambda \rightarrow m_0=M_s$ where $M_s$ is the string scale, which is correct since the degeneracy at each 
%level is so large that the species scale coincides with the string scale.
In these singular limits  the appropriate QG scale is in general not the Planck scale but rather the {\it Species Scale},
\cite{Dvali:2007hz,Dvali:2007wp,Dvali:2008ec}, which 
may  be much lower than the Planck scale depending on the
number of states   $N$ with mass below $\Lambda_{QG}$. 
In a perturbative definition for the Species Scale in d-dimensions, it  is related to
the number of species in a tower  $N$ by
\beq 
\Lambda ^{d-2}\ \simeq \ \frac {M_{\rm P}^{d-2}} {N} \ .
\label{species}
\eeq
Combining this expression with eq.(\ref{torre}) one can check that the species scale associated to a a given tower mass $m_t$
is given by 
\beq
\Lambda _t\ \simeq \  m_{t0}^{p/(d-2+p)}
\eeq
in Planck units. 
This corresponds to a particular direction in moduli space, and along a specific tower of states $m_{tn}$. This is thus a {\it local} 
definition of Species Scale. Note that the number of species depends on the moduli, $N(\phi)$, since the masses in the tower are 
field dependent themselves and so is the number of states below the cut-off $\Lambda_t$. 
 More generally, it is important to realise that the species scale is a {\it  function of all the moduli of the theory}
 \cite{vandeHeisteeg:2022btw,vandeHeisteeg:2023ubh,vandeHeisteeg:2023dlw,Castellano:2023aum,Aoufia:2025ppe},
 giving different results depending on the 
 moduli direction.  In particular the species scale should be also defined in the bulk of moduli space, not associated to any tower.
 Clearly, it would be very informative to have a closed expression for $\Lambda$ as a function of the all moduli for any given string vacua.

It has been recently realised \cite{vandeHeisteeg:2022btw,vandeHeisteeg:2023ubh,vandeHeisteeg:2023dlw,Castellano:2023aum}
that in string vacua with enough SUSY generators one can obtain in some cases explicit expressions for the species scale. 
Thus if  the species scale $\Lambda$ is a UV cut-off,  higher dimensional operators in the theory should appear
suppressed by powers of $\Lambda$. Thus an alternative definition to eq.(\ref{species}) could be obtained by identifying the 
Wilson coefficients multiplying such higher dimensional operators.
Concretely, one can consider BPS protected higher derivative gravitational operators of the general form
\beq
S_{\rm BPS} \ =\ \frac {1}{2\kappa _d^2}
\int d^d x \sqrt{-g} N {\cal F}_n^{(d)} \frac {{\cal R}^{2n} }{M_{p;d}^{4n-2}}\ ,
\label{R2n}
\eeq
with $n=2$ in the case of 32 and 16 supercharges, and $n=1$ for the case of 8 supercharges,
like 4d ${\cal N}=2$ CY compactifications of Type II string theory.  One expects the mass scale
controlling these operators to be given by the cut-off of the theory so that one has
\beq
\Lambda  \ \simeq \ M_{{\rm p}}^{(d)}\, {\cal F}_n^{-\frac {1}{4n-2}} \ .
\eeq 
Thus one can get a field dependent expression over all moduli space if one is able to compute the Wilson
coefficients  ${\cal F}_n^{(d)}$. This is the case  e.g of maximal  sugra  from Type IIA and Type IIB 
and their toroidal compactifications \cite{Green:2010kv,Green:2010wi,Green:1999pv,Pioline:1998mn,Green:1999pu,Green:2005ba}.
As an example the Wilson coefficient ${\cal F}_2$ has been computed from
the scattering amplitude involving four  gravitons in Type IIB with the general result.
\beq
S_{R^4}^{\text{10d}}= \frac{1}{\ell_{10}^2} \int d^{10}x \sqrt{-g}\, E_{3/2}^{sl_2} (\tau, \bar \tau)\, t_8 t_8 R^4\, .
\label{eq:10dR^4IIB}
\eeq
Here $t_8$ are standard helicity tensors and $E_{3/2}^{sl_2}$ is the non-holomorphic Eisenstein-Maas form of weight 3/2
corresponding to  the $SL(2,{\bf Z})$ duality symmetry of the Type IIB complex dilaton $\tau$ .  One can then write for the 
species scale $\Lambda_{IIB}^6=1/E_{3/2}^{sl_2}(\tau,{\overline \tau})$ and check that , for large $Im\tau$, one gets $\Lambda_{IIB}\simeq M_s$.
One has analogous expressions (though slightly more cumbersome) for the maximal SUSY theories obtained upon toroidal
compactification of $M$-theory \cite{Green:2010kv,Green:2010wi,Green:1999pv,Pioline:1998mn,Green:1999pu,Green:2005ba}.

 In the  case of 4d ${\cal N}=2$ theories obtained from  10d Type II 
compactifications on a CY manifold  one has
\beq
S_{{\cal N}=2} \ =\ \frac {1}{\kappa_p} \int  d^4x \ N\ 
\frac {({\cal R}\wedge {\cal R})}{M_{\rm P}^ 2} \ ,
\eeq
with $N=\Lambda^{-2}$.  Here $N$ will be a function in Type IIA of the $h_{11}$ vector moduli and $(h_{21}+1)$ hypermultiplets 
(the opposite in Type IIB).  It turns out that the dependence on vector moduli may be explicitly computed in terms of topological string theory \cite{Bershadsky:1993ta}.
In particular it was argued in  \cite{vandeHeisteeg:2022btw,vandeHeisteeg:2023ubh}
that  $N$ may be obtained from the genus-one free energy of topological strings propagating in the CY.
Thus in a class of theories the species scale (or at least its dependence on vector moduli) may be computed exactly, including perturbative and non-perturbative effects.

%An example of this is \cite{Bershadsky:1993ta,vandeHeisteeg:2022btw,vandeHeisteeg:2023ubh}
%the species scale for theories including a torus, like the Enriques CY which is $K3\times T^2/{\mathbb{Z}_2}$, see
%eq.(\ref{N-toro})  in section (\ref{section5}).
%One important general property of the species scale is that it contains all duality invariances in the theory.  An example of this is the
%SL$(2,\mathbb{Z})$ modular invariance associated to the $T^2$ in those examples. The associated species scale is modular invariant. This 
%is also known to happen in higher-dimensional string vacua like maximal sugra toroidally compactified 
%\cite{Green:1997tv,Green:2010kv,Green:2010wi}.
%This modular invariance will be an important guide in order to try to obtain information on the form of the effective potential in the following sections.

\section{Laplacians in various dimensions and the Swampland}

The first topic in my talk at  the Corfu School is based on work done in collaboration with
Christian Aoufia and Alberto Castellano in ref.\cite{Aoufia:2025ppe}.

In view of  the SDC,  which implies exponentially decaying towers of masses when moduli go to infinite distance, 
the same should apply to  the species scale along the moduli direction determining the tower,  so that one
expects a behaviour of the form $\Lambda_t\simeq \Lambda_{0t} exp(-\lambda_\Lambda \Delta \phi)$, with a rate $\lambda_\Lambda$
of order one. In fact it has been verified in many  string compactification examples that this rate is given asymptotically by
\cite{Calderon-Infante:2023ler,Calderon-Infante:2020dhm,vandeHeisteeg:2022btw,vandeHeisteeg:2023ubh}
\beq
|\lambda_\Lambda|^2 \ =\ g^{ij}\left(\frac {\partial_i\Lambda}{\Lambda}\right)
\left(\frac {\partial_j\Lambda}{\Lambda}\right) \ =\ \frac {p}{(d-2+p)(d-2)} \ .
\label{asymp}
\eeq
Thus for a single KK tower one has $p=1$ and $|\lambda_\Lambda|^2=1/(d-1)(d-2)$ whereas for a string $p=\infty$ and 
$|\lambda_\Lambda|^2=1/(d-2)$.  Indeed these expressions and the SDC itself has been tested in multitude of 
string examples,  but no clear-cut microscopic  explanation exists for the origin of this exponential dumping.

In trying to find out what is the origin of this behaviour,   a direction which seems worth exploring 
is to study the general 
properties of the species scale as a function of all the moduli of a given classical vacuum.
Following this purpose we studied some differential properties of the species scales in ref.\cite{Aoufia:2025ppe}.
Using as species scales those derived from the Wilson coefficients in the higher curvature operators in eq.(\ref{R2n}),
we found that
the Wilson coefficients are eigenfunctions of 
a   'Laplace-like' differential equations  in $d$ dimensions of the form
\beq\label{eq:eigenvalueeqintro}
\cD^2_{\bf {\cal M}} {\cF}^{(d)}_n   = \eta_d\,  {\cF}^{(d)}_n\, .
\eeq
Here $\eta_d$ are constant eigenvalues and ${\cal D}^2_{\bf {\cal M}} $ is a
second-order, elliptic differential operator
defined on the moduli space of the theory. In most examples this operator turns out to be just the Laplace-Beltrami  operator 
$\cD^2_{\bf {\cal M}}=\Delta_{\bf {\cal M}}$. This happens both for $n=2$ and $n=1$ in eq.(\ref{R2n}), for theories with 32,16 and 8 SUSY generators respectively.
Table \ref{tab:summary} summarizes the different operators an eigenvalues for different cases that we find, see ref.\cite{Aoufia:2025ppe}.
\begin{table}[]\renewcommand{\arraystretch}{1.4}
    \centering
    \begin{tabular}{|c|c|c|c|} \hline
        \textbf{Theory} & \textbf{Coefficient} & \textbf{Operator} & \textbf{Eigenvalue}  \\ \hline
         Maximal supergravity* (10d, 9d, 8d) & $\cR^{4}$ & $\Delta_\cM$ & $ 3\, \frac {(11-d)(d-8)}{d-2}$ \\ \hline
         Type IIA & $\cR^{4}$ & $\partial_\phi^2 + \partial_\phi$ & $ \frac34$  \\ \hline
         Half-maximal supergravity (10d, 9d)  &$\cR^{4}$ &  ref.\cite{Aoufia:2025ppe} & $ 3\, \frac {(11-d)(d-8)}{d-2}$ \\ \hline
         6d $\cN = (1,0)$ ($T$ tensor multiplets) & $\cR^{2}$ & $\Delta_{\cM}$ & $T$ \\ \hline
         5d $\cN = 2$ ($n_V$ vector multiplets) & $\cR^{2}$ & ref.\cite{Aoufia:2025ppe} & $\frac16 n_V$ \\ \hline
         4d $\cN =2$ &  $\cR^{2}$ & $\Delta_{\cM}$ & 0 \\ \hline    
    \end{tabular}
    \caption{\small Summary of the various operators, Wilson coefficients and associated eigenvalues, see ref.\cite{Aoufia:2025ppe}. }
    \label{tab:summary}
\end{table}
The fact that the Wilson coefficients for maximal sugra are  solutions of a Laplace-like  equation was already known
\cite{Green:2010kv, Green:2010wi, Green:1999pv, Pioline:1998mn, Green:1999pu, Green:2005ba}.
In \cite{Aoufia:2025ppe}  we reconsider this fact from the Swampland point of
view and show how, in d = 10,9,8, solving a Laplace equation imposes non-trivial restrictions
on the species scale behaviour. We further argue that this property is also satisfied in settings
with less supersymmetry. In particular, for the  ${\cal R}^4$-operator in minimal supergravity
theories in d = 10, 9,  (16 supercharges) and on the leading  ${\cal R}^2$-term in setups with 8 supercharges in d = 6, 5, 4.

The fact that the Wilson coefficients (related to the species scale) obey these Laplace equations 
implies certain consistency conditions on the species scale asymptotic behaviour like   $\lambda_\Lambda^2 \leq 1/(d-2)$,
proposed as a conjecture \cite{vandeHeisteeg:2022btw,vandeHeisteeg:2023ubh}.  Let us consider as an example the Type IIB case already mentioned. In this case
from Table \ref{tab:summary} one sees that the relevant Laplace equation is
\beq
\Delta_{sl_2} {\cal F}(\tau,{\overline \tau}) \ = \ \frac {3}{4} {\cal F}(\tau,{\overline \tau})  \ ,
\eeq
with $\Delta_{sl_2}=\tau_2^2(\partial_{\tau_1}^2+\partial_{\tau_2}^2)$.
The idea now is to consider an ansatz for the large $\tau_2\rightarrow \infty$ limit, ${\cal F}\ \sim \ \tau_2^\lambda $. 
Then the eigenvalue equation yields 
\beq
\lambda(\lambda-1) \ =\ \frac {3}{4} \ \longrightarrow \ \  \lambda\ =\ 3/2, -1/2  .
\eeq
This matches with the leading 'constant'  terms in the Fourier expansion of the $E_{3/2}$ Eisenstein form mentioned above
\beq
E_{3/2}^{sl_2} \ =\ 2 \zeta(3) \tau_2^{3/2} \ +\ 4\zeta(2) \tau_2^{-1/2}\ +\ {\cal O}(e^{-2\pi \tau_2} ) \ .
\eeq
For  large dilaton  the Laplace equation then predicts
\beq
\Lambda_{IIB}\ \simeq \ \frac {M_{p10}}{ {\cal F}^{1/6}}\ \simeq \ \frac {M_{p10}}{\tau_2^4}\ \simeq \  M_s,
\eeq
i.e., as we said, the species scale is given by the string scale $M_s$  in IIB. In canonical basis for $\tau_2$ one recovers the expected behaviour
from the SDC,  $\Lambda\sim exp(-\phi/\sqrt{d-2})$. Thus the SDC in this example may be considered a consequence of the above Laplace equation.
The constraints from the Laplace equations for Type IIA and maximal sugra  in $d=9,10$ may be also analysed. Thus e.g. one can consider 
M-theory compactification on $T^2$. Now one has more moduli, one complex $SL(2,{\bf Z})$ modulus $\tau$ and a real scalar ${\cal V}$, and the eigenvalue is
$6/7$. In terms of canonical variables $U,{\hat \tau}$ one can write  as an ansatz ${\cal F}\sim e^{Ux+{\hat \tau}y}$ and impose the Laplace equation .
We obtain a constraint in moduli $x,y$ space
\beq
\frac {1}{2} x\left(x\ +\ \frac {2}{\sqrt{14}} \right) \ +\ \frac {1}{2} y\left( \ y\ -\ \sqrt{2}\right)\ =\ \frac {6}{7} \ .
\eeq
Now the space of solutions is a continuous curve.
\begin{figure}
    \centering
    \includegraphics[width=0.55\linewidth]{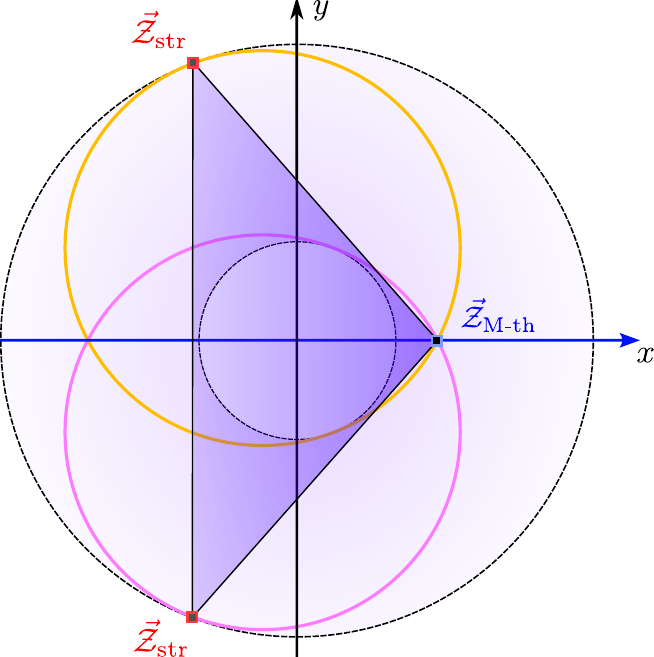}
    \caption{\small Laplace circles and species polytope for the case of maximal supergravity in $d=9$. The outer dashed circle is 
    the upper bound  $1/\sqrt{(d-2)}$, see \cite{Aoufia:2025ppe} for details.}
    \label{circulos}
\end{figure}
Converting the condition on ${\cal F}$  to a condition on the species scale $\Lambda$ one gets the two circles  in yellow and pink in
fig.(\ref{circulos}). In this figure the species scale convex-hull obtained from the vectors ${\mathcal Z}^i=-g^{ij}(\partial_j log(\Lambda))$ is 
displayed as a triangle. It is easy to check that any solution in the circles obeys the bound $|\vec {{\mathcal Z}}|^2\leq 1/(d-2)=1/8$,
giving an explanation for this bound in terms of the Laplace equation.
This bound is saturated at the upper and lower vertices, which correspond to a string tower and its $SL(2,{\bf Z})$ dual. At the other vertex 
lying on the circle intersection is the full decompactification up to 11d M-theory limit. 

A slightly more complicated structure is obtained for maximal sugra in 8 dimensions in which the loci of solutions are now spherical surfaces
which again very much constrain the form of the convex hull  polytope. A Laplace equation also  appears for vacua with less SUSY generators.
In particular the 10d Heterotic strings with 16 SUSY generators, which have the dilaton as a single modulus. It turns out that for the 
$SO(32)$ case the Laplace equation is the same as in Type IIB setting the $C_0$ axion to zero. On the other hand the $E_8\times E_8$ heterotic
has a Laplace equation similar to that in Type IIA. For the 9d reduction with $SO(16)\times SO(16)$ group , the equation is in the same way related to
the maximal sugra 9d example.

In theories with only 8 supercharges  also eigenvalue Laplace-like equations exist. In all cases the dependence on vector(tensor) moduli of the
Wilson coefficients is asymptotically linear.
Thus in the case of 6d  ${\cal N}=(1,0)$  supergravity vacua the moduli space is 
spanned by scalars in the tensor multiplets, and it is a slice in the coset $SO(1,n_T)/SO(n_T)$. The Wilson coefficient of 
${\cal R}^2$ terms obeys a Laplace-Beltrami eigenvalue equation of the simple form
\beq
\Delta_{{\cal M}} {\cal F} \ =\ n_T\ {\cal F}\ ,
\eeq
wih $n_T$ the number of tensor multiplets. In the 5d case , like M-theory on a CY, there is an eliptic second  order operator ${\cal D}^2$ which differs from the
Laplace Beltrami one \cite{Aoufia:2025ppe}. Modulo this modification the structure is somewhat similar with
\beq
{\cal D}^2_{{\cal M}} {\cal F} \ =\  \frac {1}{6} n_V\ {\cal F}\ ,
\eeq
with $n_V$ the number of vector multiplets.  In the case of 4d, like  Type IIA(B) on a CY, the ${\cal R}^2$ operator is marginal, and infrared field theory 
corrections are expected \footnote{This is also true for $d=8$ in maximal sugra case for the ${\cal R}^4$ operators. }. These infrared contributions play a 
negligible role if we are interested in  computing the number of species so we ignore them in what follows. Then the associated elliptic 
equation is the Laplace-Beltrami equation with
\beq
\Delta_{4d} {\cal F} \ =\  0 \ ;\  \Delta_{4d} = 2 g^{a{\bar b}} \partial_a\partial_{\bar b} \ ,
\eeq
with $g_{a{\bar b }}$ the metric in vector moduli space.
This equation  is indeed satisfied in string theory  examples up to the holomorphic anomaly related
to IR corrections  from massless modes within the theory that we mentioned.
The condition, together  with the assumption of axion independence at large vector moduli,   is enough to determine the general form of the Wilson coefficient which
must be linear in the  saxions ${\cal F}_{4d}\sim c_{2,a}t^a$. 
This is in agreement with   known string theory results in which the $c_{2,a}$ are related to the second Chern class of the CY. One thus has
\beq
\Lambda_{4d} \ \sim \ \frac {M_{p,4}}{\sqrt {c_{2,a}t^a}} \ .
\eeq
From this general expression, after going to canonical local frames, one can then derive the asymptotic behaviour for the 
species scale in  different infinite distance limit. 
We also find a relation between the above Laplace constraint in 4d and certain versions of the 
Scalar Weak Gravity  \cite{Palti:2017elp,Lee:2018spm,Gonzalo:2019gjp,Gonzalo:2020kke,DallAgata:2020ino,Andriot:2020lea,Etheredge:2022opl,Benakli:2022shq,Dudas:2023mmr,Etheredge:2023usk}, in particular the one discussed in \cite{Gonzalo:2020kke}.

We have seen that not always the relevant second order operator is just the Laplace-Beltrami operator. 
In those special cases one can see that their naive continuous isometries from the form of the kinetic  $\sigma$-model are
not respected by some terms in the two-derivative action. A
prototypical example of this is 10d Type IIA, where the naive shift symmetry associated
to the dilaton  $\phi \rightarrow \phi + \lambda$ is broken by  $e^\phi$ couplings with other sectors. 
In string theory examples dualities and dimensional reduction allow us to figure out the form of the 
asociated operator ${\cal D}^2$ in those specific cases.

As a summary, the fact that the Wilson coefficients obey Laplace-like
equations strongly constrain the asymptotic behaviour at infinite distance limits of the species scale, 
providing in some cases  for an explanation for exponential behaviour ( SDC) and  conjectured bounds on the rates. 

The question is: why these Laplace equations are obeyed to begin with? What is the origin of this very particular behaviour 
of  the Wilson coefficients, directly connected to the species scale? We do not have yet an answer to this question. 
One may speculate noting that  the Laplace equation here is reminiscent of dissipative systems in which the time evolution of a scalar function $S$ is governed by an equation
('2nd Ficks law') $\partial_tS = c\Delta S$. If $S$ obeys a Laplace equation with non-vanishing eigenvalue this leads for an exponential behaviour 
for the quantity $S$. This possible  connection is just intriguing but,  more generally,  it should be worth exploring what underlies these Laplace 
conditions present in string vacua.

\section{The species scale as a field dependent UV cut-off  and moduli fixing}

This second topic that I covered in the Corfu 2025 conference was an application of the species scale as a {\it moduli-dependent}  UV cut-off in 4d  EFT's,
and is based on work done in collaboration with Gonzalo F. Casas in ref.\cite{Casas:2025bvk}.

\begin{figure}[h!]
            \hspace{2em}
            \includegraphics[width=0.9\linewidth]{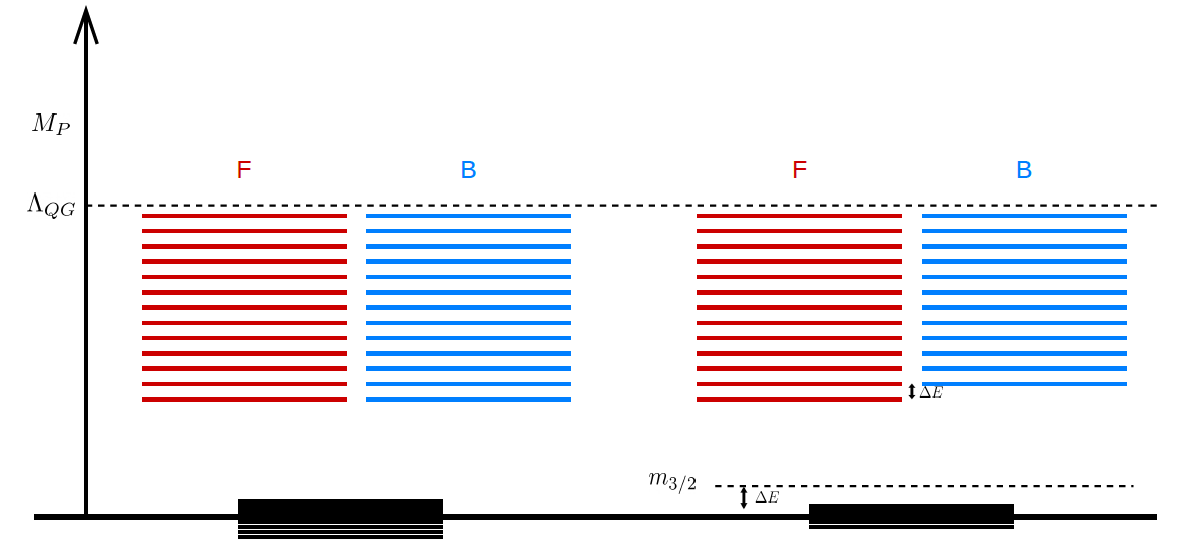}
            
            \caption{The different towers of states and the light states with unbroken SUSY are depicted on the left of the picture. On the right, SUSY is spontaneously broken, the gap between states is given by the gravitino mass, and there are generally fewer light states in the theory.}
            \label{structurescales}
        \end{figure}

We consider in particular 4d theories with ${\cal N}=1$ which might have some application to the issue of moduli fixing in realistic string compactifications. 
We will be typically interested in the EFT in a theory in which some modulus is large so that one expects a tower of (KK or string) states coming down.  Schematically, the
spectrum of states (before SUSY breaking) is as in the left picture in fig.\ref{structurescales},  with boson-fermion degeneracy. There is some massless states and
a tower (say, in the KK case) whose lightest state has mass $M_{KK}$.  We want to study the moduli  dependent  EFT below the $M_{KK}$ scale   taking 
into account that 1)  we take as UV cut-off the
species scale and 2) take  into account that the cut-off itself is moduli dependent. 

To be more specific we have in mind the case of 4d ${\cal N}=1$ Type IIB orientifold models. We consider 
classical non-SUSY vacua in Minkowski space in
4D, obtained as no-scale solutions. The classical examples are GKP-like configurations
in Type IIB theory and generalizations.
These kinds of
theories have chiral multiplets $Y_a$  which explicitly appear in the flux superpotential
$W_{flux}(Ya)$,  with  $D^a W_{flux} = 0$. The rest of the moduli denoted  $Z_i$ may contribute to the
SUSY-breaking and their auxiliary fields verify a no-scale condition  $g_{i{\bar i}}|F ^i|^2 = 3m_{3/2}^2$
leading to Minkowski vacua.  The $Z_i$ remain at this level as classical moduli, and we will
call them no-scale moduli. Accordingly, in the Type IIB GKP case, the  $Y_a(Z_i)$ will be
complex structure + complex dilaton (Kahler moduli) respectively.

\subsection{The one-loop potential from a tower}

\begin{figure}[h!]
            \centering
            \includegraphics[width=0.4\linewidth]{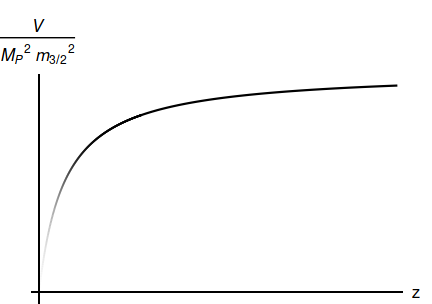}  \ \ \ \ 
           \ \ \ \ \ \ \  \includegraphics[width=0.4\linewidth]{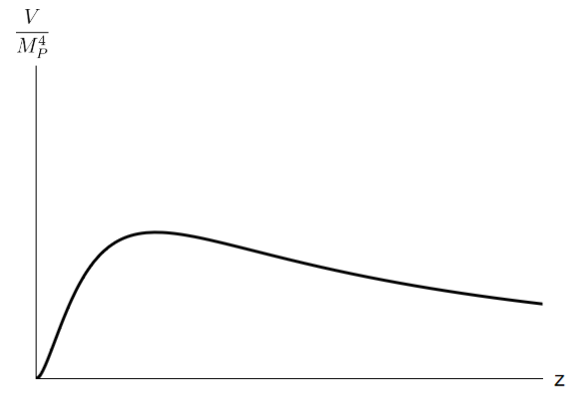}
            \caption{1) Qualitative behaviour of equation  $V_{1-loop}/g^2m_{3/2}^2$  from the first EFT computation in eq.(\ref{V1loop}).
            2) 
            Qualitative behaviour of the full 1-loop potential from eq.(\ref{poten1}), extrapolated down to a {\it  desert point}.}
            \label{simple}
        \end{figure}

The no-scale moduli are classically  massless but one expects they will get a non-trivial potential due to one-loop corrections.
The one-loop potential in a theory with a cut-off $\Lambda$
has the form
%\cite{PhysRevD.7.2887,PhysRevD.7.1888,Schwartz_2013}
%
\begin{equation}
    V_{\rm total}^{\rm eff} = \frac{1}{(8\pi)^2}\left(-\text{Str}\mathcal{M}^0\Lambda^4 + 2\,{\rm Str}\mathcal{M}^2\Lambda^2 - 2\, {\rm Str}\mathcal{M}^4 \log\left(\frac{\Lambda}{\mathcal{M}}\right)\right) \ ,
    \label{potencial}
\end{equation}
where we have defined the supertraces
\begin{equation}
    {\rm Str}\mathcal{M}^a = \sum_k (-1)^{2j_k}(2j_k + 1)( m_k)^a,\quad a=0,2,4,
    \label{supertraza}
\end{equation}
and $j_k$ is the spin of each particle $k$.  Note that in a SUSY theory one has Str$\mathcal{M}^0=0$, so that the leading term will be the second one.
The following term in ${\rm Str}{\cal M}^4$ may be shown to be subleading \cite{Casas:2025bvk}. In our case the one-loop potential will have two leading contributions from the light 
states and from the tower of massive states, i.e.
\beq 
V_{1-loop}^{eff}\ \simeq \ \frac {2}{(8\pi)^2}\left({\cal S}tr  {\cal M}^2_{light} \ +\ {\cal S}tr {\cal M}^2_{tower} \right) \ \Lambda^2 \ .
\eeq
Concerning the first piece, if the massless moduli have hyperbolic metrics, as is generically the case, 
one can compute  the supertrace by using the general ${\cal N}=1$ supergravity expression 
\cite{Ferrara:2016ntj}:
\beq
{\rm Str}{\cal M}^2\ =\  2(N_0-1)m_{3/2}^2 \ +\ 2e^{K}{\cal R}_{i{\bar j}}(D^iW)({\overline D}^{\bar j}{\overline W}) \  .
\label{ferrara}
\eeq
Here $N_0$ is the number of chiral multiples, and ${\cal R}_{i{\bar j}}$ is the Ricci tensor in moduli space. 
One then finds \cite{Casas:2025bvk}
\beq
{\rm Str}{\cal M}_{{\rm light}}^2\ =\ -4 m_{3/2}^2\,
 \left( N_0  + 2\right) \ .
 \eeq
Concerning the contribution of the massive tower, which  is compossed of chiral multiplets,
after SUSY breaking their bosonic components get a  shifted mass$^2$ so that
$m_B^2\rightarrow m_B^2+m_{3/2}^2$.  Then one can sum over the tower to obtain
\begin{equation}
    {\rm Str}\mathcal{M}^2\Lambda^2 \simeq  4\Lambda^2\sum_n^N( n^{2/p} m_0^2 + m_{3/2}^2 - n^{2/p} m_0^2 ) \simeq  4\Lambda^2 N m_{3/2}^2 \simeq 4m_{3/2}^2 M_{\rm P}^2 \ ,
\end{equation}\newline
where $m_0$ is the mass of the lightest state in the tower and $p$ is the density parameter described above,
and we have also used the defining equation $\Lambda^2\simeq M_{\rm P}^2/N$.  Interestingly, this contribution is independent of the
UV cut-off $\Lambda$. It is also independent of the density parameter $p$ of the tower. Thus, it applies not only for KK towers but also for the string case $p=\infty$.
Altogether one gets a final structure for the 1-loop potential
\beq 
V_{1-\rm{loop}} \  \simeq  \ \frac {g^2m_{3/2}^2M_{\rm P}^2}{(8\pi)^2}   \left( c \ - \ \frac {\eta}{N(z_i,{\bar z}_{\bar i}) }\right)  \  ; \ \ \eta = 8\left(N_0+2\right)\ ,
\label{V1loop}
\eeq
with $c$ a positive constant of order one. When performing a string loop computation, in the Einstein frame there is a loop-counting factor $g^2=e^{2\phi_4}$
multiplying which is here included.
The first term comes from the tower, while the second arises from the light modes.
Note that for fixed gravitino mass the potential is minimised as $N\rightarrow {\cal O}(1)$, which in terms
of moduli corresponds to $z_i\sim {\cal O}(1)$, with the species scale $\Lambda$ approaching its 
maximum value close (below) $M_{\rm P}$. This is reasonable, $N$ counts the number of states in the tower which,   is known to decrease 
as the moduli vev decreases (recall that $N$ has a linear behaviour with the moduli). Thus the potential should decrease as the modulus decreases.
This is schematically depicted in  Fig. 
 \ref{simple}. This shows the behaviour  of $V_{1-loop}$ for constant $g^2$ and $m_{3/2}$, showing a trend  towards the origin as moduli decreases 
 and flattens off for large modulus. In fact in general both $g^2$ and $m_{3/2}^2$  decrease for large Kahler moduli in string compactifications, so that the full
 potential will show a local maximum at intermediate values and asymptotically decreases due to the $g^2m_{3/2}^2$ factor.
 Thus e.g. 
 the gravitino mass depends on the no-scale saxions like $m_{3/2}^2\ \sim \ 1/u^r$,  and $e^{2\phi_4}\sim  1/\sqrt{su^r}$ for a positive $r\leq 3$.

The above computation is performed for large moduli and shows a tendency for the modulus to be fixed dynamically  at small values. 
This is interesting since e.g. in the context of Type IIB orientifolds this would mean that
many  Kahler moduli could be fixed at small values without the need for non-perturbative effects. On the other hand the 
computation does not give direct information about the behaviour in the bulk, but just a general qualitative tendence.

\subsection{ The one-loop potential and the desert points}

 We will try now to obtain some information about the behaviour in the bulk by using known properties of the
species scale. In particular one can compute one-loop corrections to the metric $g_{i{\bar i}}$ of the massless field due to loops from  the tower states
summed up to the species scale. This type of computation has been performed recently 
\cite{Grimm:2018ohb,Heidenreich:2017sim,Heidenreich:2018kpg,Castellano:2022bvr,Castellano:2023qhp,Blumenhagen:2023yws,Casas:2024ttx}
in the context of the 'emergence proposal' so we will not
give many details. Furthermore we will not assume here any form of the 'emergence proposal', but
simply use the one-loop computations as detailed in e.g.\cite{Castellano:2022bvr}.
The $Z_i$ moduli couple to massive modes through trilinear couplings
$\partial_im_{ni} = n\partial_im_{0i}$, with $m_{0i}$ the KK scale.
Then a one-loop graph would give rise to a correction
\beq
%\begin{align}
\delta g_{i{\overline i}}  \simeq  \frac {4\,g^2}{(8\pi)^2} \sum_n  n^2  (\partial_im_{0i})(\partial_{\bar i}m_{0i}) 
\simeq  \frac {4\,g^2}{(8\pi)^2} N^3  (\partial_im_{0i})(\partial_{\bar i}m_{0i})\label{correctiondG}
\simeq    \frac {4\,g^2}{(8\pi)^2}  \frac {(\partial_im_{0i})(\partial_{\bar i}m_{0i}) } {m_{0i}^2}\,M_{\rm P}^2.
%\notag,
%\end{align}
\eeq
We have approximated the sum by an integral, and in the last step used $\Lambda=Nm_{0i}$ and $M_{\rm P}^2= N\Lambda^2$. This structure is the same for all $p$. 
Given a tower saturating the species scale one can use $m_{0i}= \Lambda^{(d-2+p)/p}$ as discussed above, and write \eqref{correctiondG} in terms of the species scale 
as
\beq
\delta g_{i{\overline i}} \ \simeq \ \frac {4\,g^2 }{(8\pi)^2}\frac {(\partial_i\Lambda )(\partial_{\bar i}\Lambda)} {\Lambda^2}\,M_{\rm P}^2\ \ .
\label{metrica}
\eeq
In superspace notation, the correction to the kinetic term may be written as a D-term,
\beq
\delta{\cal L}_{kin} \ =\ \left[ \delta g_{i{\overline i}} Z^i   {\overline  Z}^{\overline i} \right]_D\ \simeq
 \  \left[ \frac {4\,g^2 }{(8\pi)^2} \  \frac {(\partial_i\Lambda)(\partial_{\bar i}\Lambda)} {\Lambda^2}\ 
\   Z^i {\overline  Z}^{\overline i}\right]_D\, M_{\rm P}^2 .
\label{lagrange}
\eeq
From $(Z^i {\overline  Z}^{\overline i})_D$   one recovers the one-loop correction to the modulus
kinetic term metric. However, setting   $Z^i = \theta^2F^i$ , with $F^i$ the ${\cal N}=1$ auxiliary field,  one recovers a one-loop contribution to the vacuum energy 
\beq
\delta V \ \simeq \ \frac {4\,g^2\,m_{3/2}^2 M_{\rm P}^2 }{(8\pi)^2}  \ g^{i{\overline i}} \frac {(\partial_i \Lambda ) } {\Lambda}  \frac {(\partial_{\overline i} \Lambda)}  {\Lambda }   \ ,
\label{poten1}
\eeq
where we have used that $F_iF^{\bar{i}}=3\,m_{3/2}^2$ and  $g^{i\bar{i}}$ is the tree level metric. Recalling eq.(\ref{asymp}) one sees that for large modulus one has
a behaviour for the potential $\delta V\sim g^2m_{3/2}^2M_p^2 /(8\pi)^2$, in agreement with the previous computation in eq.(\ref{V1loop}).

Although computed for the large modulus limit, we would like to argue that this expression has the correct structure so that it may be extrapolated 
to the bulk. A first argument is the intuition that the scalar potential should vanish at the {\it desert points} . The reason for that is that close to those points there are 
no states (like e.g. BPS states in ${\cal N}=2$ subsectors of the theory) to sum over, no states below the species scale which is now of order the Planck scale,
so no one-loop contribution. The above loop potential verifies this since, indeed, by definition $\partial_i\Lambda = 0$ at the desert points. 
Secondly, in models with duality symmetries like e.g. $SL(2,{\bf Z})$, modular invariance restricts the form of the potential and connects the large modulus limit 
with the behaviour in the bulk. We will momentarily see an example in which the species scale is known,
and the potential, as computed from eq.(\ref{poten1}) , is proportional to the modular invariant
function  $u^2|{\tilde G}_2(U,{\overline U})|^2$, where ${\tilde G}_2$ is the (non-holomorphic)  weight-2 Eisenstein modular form.
Strictly speaking we only know eq.(\ref{poten1}) to be correct for large $u$, in
which limit one has  $u^2|{\tilde G}_2(U,{\overline U})|^2\rightarrow u^2$.
Thus the idea is that modular invariance dictates
that $u^2$  is completed in the bulk as  $u^2\rightarrow u^2|{\tilde G}_2(U,{\overline U})|^2$.
 But this corresponds to extending
the validity of the computation eq.(\ref{poten1}) to the bulk in moduli space.  Not doing this would lead to a non modular invariant scalar potential.
By the way, this extrapolation
is the analogue of the replacement  $t\rightarrow -log(t|\eta(T)|^4$  in the heterotic one-loop threshold correction example discussed below. 

Note that the potential above is positive definite and have minima at the desert points. Thus the general structure, including the expected large modulus 
suppression by the factor $g^2m_{3/2}^2$, is depicted in the plot  in the right in fig.(\ref{simple}). In order to
see that this  computation also yields a negative correction going like  $\sim -1/N$ 
we have to be a bit more specific about the form of the species scale and the class of models we are discussing. 
To be more explicit, let us consider first the class of CY Type II compactifications leading to ${\cal N}=2$ theories in 4d. An additional
orientifold projection may lead to ${\cal N}=1$  (with some ${\cal N}=2$ subsector), which one can expect to inherit much of the results in the discussion.
In those ${\cal N}=2$ models one can compute the function $N$ in terms of the vector moduli using tools of Topological Field Theory 
\cite{Bershadsky:1993ta,vandeHeisteeg:2022btw,vandeHeisteeg:2023ubh}.
The details depend on the particular type of infinite limit, e.g., large volume, emergent string limit, or 6d decompactification, as well as the Hodge 
numbers of the CY compactification. For all relevant infinite distances, parametrizing the limit in terms of a single large modulus  $s$, one gets the
structure \cite{vandeHeisteeg:2023ubh}
\beq 
N\ =\ \frac {2\pi c_2}{12} s \ -\ \beta \log(s) 	 \ +\ \tilde{N_0} \ + \   {\cal O}(s^{-1}) \ .\label{expansion}
\eeq
Here, $c_2$ is related to  the second Chern class of the CY. The value of $\beta$ depends on IR details,
in particular the Hodge numbers $h_{11},h_{21}$, and is in general a positive number, see ref.\cite{vandeHeisteeg:2023ubh}.
The constant $\tilde{N_0} $ denotes the number of states at the desert point.
Inserting \eqref{expansion} in \eqref{poten1}, up to subleading logarithmic terms 
takes the form,
              \beq
              V_{1-\rm{loop}}   \ \simeq \  \frac {g^2\,m_{3/2}^2 M_{\rm P}^2}{(8\pi)^2}
              \left( 1\ -\  \frac {\eta}{s}\right)^2 ,\ \quad \eta=\frac {12 \tilde{N_0} }{2\pi c_2 }.
              \label{poton}
              \eeq
              Thus, indeed, $V_{1-\rm{loop}}$ has the behaviour of a plateau and a negative one-loop correction of the form $\sim (-1)/N$, as indicated by the
              previous EFT computation. It also captures a similar constant $\eta$, which, as in the previous case, depends on the number of states. 
In  CY compactifications with a 2-torus factor, like in the Enriques CY or toroidal 
               orientifolds, the form of the species scale depending on vector moduli 
                may be obtained from the Topological String Theory computation in \cite{Bershadsky:1993ta}.
                              One has  %
                 \begin{equation}
    N = -6 \log(2u |\eta(U)|^4 ) + \tilde{N_0}\ ,
    \label{N-toro}
\end{equation}
with $u={\rm Im}U$ the torus modulus.  
   From here, one would then compute the one-loop potential from eq.(\ref{poten1}) and obtain 
   \beq
   \delta V \ \simeq \ \frac {2g^2\,m_{3/2}^2M_{\rm P}^2}{(8\pi)^2}  \left(\frac {3}{\pi}\right)^2 \left| \frac {u}{N} {\tilde G}_2(U,{\overline U} )\right|^2 \ 
   \label{inflapot}
   \eeq
               Here      ${\tilde G}_2$ is the  Eisenstein (non-holomorphic)   weight-2  modular form. It may be written in terms of the holomorphic 
         (although non-modular covariant) weight-two  Eisenstein form $G_2$ as
         \beq
         {\tilde G}_2(U) = G_2  - \frac {\pi }{u} \  ;\  G_2 \ =\ \frac { \pi^2}{3} (1 -24e^{2\pi iU}\ +\dots ) \ .
         \eeq
It is well known  that any modular invariant function like the one above has extrema at the self-dual points $ U=\{i, e^{i\pi/3}\}$, and in 
particular  ${\tilde G}$ vanishes there. One has the asymptotic behaviour
\beq
N \longrightarrow 2\pi u - 6\log(2u) + \dots \  \ ;\ \ 
G_2 \longrightarrow \frac{\pi^2}{3} + \dots  \ \ ;\ \ 
\Lambda \longrightarrow (2\pi u)^{-1/2} \,.
\eeq
Substituting into the expression for the potential, one obtains (again, up to subleading corrections):
\begin{equation}
\delta V \simeq \frac {g^2\,m_{3/2}^2 M_{\rm P}^2 }{2(8\pi)^2}\,  \left( 1 - \frac{2\tilde{N_0}}{N} + \dots \right),
\end{equation}
which has the expected structure.
We thus see that, in spite of $u\sim {\cal O}(1)$ being in  a
region of moduli space  in which corrections are large, 
 modular invariance has allowed us 
to find out that there is a minimum of vanishing energy at a self-dual point.

It is interesting to remark that in \cite{Casas:2024jbw} it was  proposed a modular invariant inflaton potential 
with  precisely the form in eq.(\ref{inflapot})  (for constant $g^2m_{3/2}^2)$). At large moduli it turned out
to have a form very similar to that of the Starobinsky potential.
This was further generalized by 
Kallosh and Linde in \cite{Kallosh:2024ymt,Kallosh:2024kgt,Kallosh:2024pat,Carrasco:2025rud} to more general classes of
modular invariant inflaton potentials. It is interesting to see that this type of inflaton potentials may admit an
interpretation as a one-loop correction in an underlying no-scale classical structure.

\subsection{Discussion}

The presence of these potentials, which generically appear (locally)  for any Kahler modulus in IIB orientifolds, may be important for the issue of 
moduli fixing. In the  KKLT \cite{Kachru:2003aw} or LVS \cite{Balasubramanian:2005zx,Conlon:2005ki} scenarios  the Kahler moduli are  typically fixed by non-perturbative effects, whereas complex dilaton 
and complex structure fields are stabilised with flux superpotentials.  Including the one-loop potentials here described some or all of the 
Kahler moduli could be stabilised  at the desert points.  On the other hand it is important to realise that for large Kahler moduli, due to the  
dilaton factor $g^2=e^{2\phi_4}$ which is volume suppressed like $\simeq g_s/Vol$, the impact of the correction here considered on the
KKLT or LVS minima is negligible. Thus  one could be in a situation in which some of the Kahler moduli are fixed in the bulk and other
are fixed through non-perturbative effects a la KKLT/LVS. These minima could also play a role in the 
generation of an axiverse as considered e.g. in \cite{Fallon:2025lvn}.

One-loop corrections in string vacua of this class and moduli stabilization have been previously discussed
at the EFT level in e.g. \cite{vonGersdorff:2005bf,Berg:2005yu,Cicoli:2008va,Antoniadis:2018hqy}. 
Explicit one loop string computations of the Kahler potential for ${\cal N}=1$ 4d orientifolds were performed 
first in \cite{Berg:2005ja,Berg:2014ama}. In particular \cite{Cicoli:2008va} includes a discussion of
the one-loop string corrections contributions understood in terms of a Coleman-Weinberg type computation. 
We do a similar CW computation but crucially include the contribution of massive e-g. KK states to the
traces, which, as we said,  gives an additional term which correctly matches the string computation 
in \cite{Berg:2014ama}.

Here we have taken the species scale $\Lambda$ as the field-dependent UV cut-off in the EFT. It may be argued that one should use instead as a cut-off the 
mass of the lightest state in a given tower, 
see e.g. \cite{Cicoli:2013swa,Burgess:2023pnk,ValeixoBento:2020gpv,Blumenhagen:2023yws,Calderon-Infante:2025ldq}
 for some discussions on this issue. This would imply considering only the contribution of the lightest (massless before SUSY breaking)
 states in the theory. However that procedure misses the potentially important {\it threshold effects} due to the massive modes which indeed 
 cannot be neglected. In particular, the one-loop vacuum energy computation at hand is just a trace over states
and is not sensitive to details of the EFT like higher dimensional operators, depends  only 
on the spectrum (light and heavy) of the theory and not on the specific form of the EFT.  It would be inconsistent to neglect the
contribution of the finite number of states, however small, which lie below the Species Scale.

To further  clarify why using the species scale as cut-off in loop computations
makes sense, let us consider an explicit well known example showing that, indeed, in order to
better match the EFT corresponding to a full string theory computation for large moduli field values,  one should sum over 
massive modes. This is the classical example of one-loop corrections to the gauge 
kinetic function in heterotic 4d orbifolds with a structure e.g. $M\times T^2$
apropriately modded to reduce to ${\cal N}=1$. 
The tree level gauge kinetic function 
has the form $f_{\rm tree}= S$, with $S$ the complex dilaton. Consider now the one-loop
 contribution coming just from light modes (the only modes 
below $m_{\rm KK}\simeq t^{-1/2}$, with $t$, the Kahler modulus of $T^2$).  One would obtain a 
contribution of order $\sim \log(t)$. However, one should  also consider in the loop 
the contribution of a tower of charged KK states with masses $m_n^2\simeq m_{KK}^2n^2/t$, summed only up to the species (string) scale. One can then compute (see e.g. \cite{Palti:2019pca,vanBeest:2021lhn,Castellano:2022bvr} for details)
\beq 
\delta {\rm Re}f_{\rm tower} \ \simeq \ \sum_n^N   (qn)^2 \ \sim \ q^2 N^3\ \simeq \ 
\frac {M_p^2}{m_{\rm KK}^2}\ \sim \ t \  ,
\eeq
where we have aproximated the sum with an integral and used $Nm_{KK}\simeq \Lambda$.
Let us emphasise that to get this result it is crucial to use as a cut-off the species scale.
On the other hand explicit string loop corrections to the gauge couplings in this setting were computed  in the classical papers \cite{Dixon:1990pc,Antoniadis:1991fh,Antoniadis:1992rq} with a result of the qualitative form
\beq 
\delta {\rm Re}f_{\rm string} \sim \ -\log(t|\eta(T)|^4) \ \sim \ t .
\eeq
For large $t$ the leading string computation is linear with $t$, matching 
the above result obtained integrating over massive modes up to the species scale. 
Had we used $m_{KK}$ as a cut-off, we would have obtained only  the $\log(t)$
subleading term. This confirms the idea that the procedure of summing over 
massive modes up to the species scale correctly captures  the full string theory 
computations for large moduli field values. 
An additional confirmation is that  the 
one-loop correction to the metric obtained  using our approach in Type II orientifolds 
correctly matches the behaviour of the one-loop string computations in ref.\cite{Berg:2014ama,Haack:2018ufg}
\footnote{Note that, irrespective of any species scale argument, 
the mere  existence of this string correction \cite{Berg:2014ama}
to the metric immediately gives rise via  the first equality in eq.(\ref{lagrange})
to the same leading correction $V_{1-loop}\sim g^2m_{3/2}^2M_p^2$ to the scalar potential at large modulus that we find.}.

As a summary we can state that no-scale moduli $Z_i$ leading asymptotically to a tower of states get induced a 1-loop potential
of the general form
\beq
\boxed {
V_{1-loop} \ \simeq \ \frac {g^2m_{3/2}^2M_p^2}{(8\pi)^2} f(Z,Z^*) \ ,
}
\eeq
where $f$ is a function such $f\rightarrow constant$ as $ImZ\rightarrow \infty$ and vanishes at desert points $Z=Z_{desert}$. Qualitatively the potential is like the rightmost curve in fig.(\ref{simple}).
The structure is
intuitively reasonable. It should be proportional to $g^2m_{3/2}^2$ since it appears at onle-loop and only after SUSY-breaking. It vanishes at the
desert points, since there are then no states below the species scale contributing. On the other hand the constant behaviour of $f$ at large 
$Im Z$ is due to the fact that the metric gets also a one-loop correction which is proportional to the tree level one, $\delta g_{ii}\propto e^{2\phi_4}g_{ij}/(8\pi)^2$. 
This may be seen from eq.(\ref{metrica}) recalling $\Lambda^2\simeq 1/ImZ$ at large moduli. 
As we said, this is consistent with the string computation results \cite{Berg:2014ama}.
This correction to the metric spoils the no-scale structure of the classical metric, and leads 
 to a modified Kahler  potential of the form $(-3+\epsilon)log(ImZ)$  at large modulus.

It is  interesting to consider this mechanism,  which can potentially fix all Kahler moduli in Type IIB no-scale orientifolds,  for
specific model building. Some preliminary steps in this direction are taken  in \cite{Casas:2025bvk} in the context of $Z2\times Z_2$ Type II orientifolds.
Although, as discussed above,  such bulk minima are expected to be present on general grounds, a limitation for a quantitative use of the mechanism is that, in the context of GKP minima,   it requires the knowledge of the dependence of
the species scale $\Lambda$ on fields that are hypermultiplets (rather than vector multiplets) in the underlying ${\cal N}=2$ theory. 
While general results 
for the ${\cal R}^2$ Wilson coefficients are known for ${\cal N}=2$ vector multiplets, hypermultiplets do not couple to ${\cal R}^2$.
 However hypermultiplet dependent Wilson coefficients may appear multiplying  other higher dimensional operators 
 (see e.g.  the case $(\partial^2S)/(ImS^2)$ in ref.(\cite{Antoniadis:1997zt})), and one could extract hypermultiplet dependent
 Wilson coefficients  (and hence species scale functions) from them. 
 
 As a general summary, the Species Scale has proved to be a useful tool to study all different asymptotic infinite limits of
 a given string vacuum simultaneously.  Crucially, it is defined over all moduli space for a given vacuum, and not only along 
 asymptotic tower directions.  Explicit field dependent Species Scale functions may be obtained in some cases from higher dimensional 
 protected operators. Interestingly, we have found that the associated Wilson coefficients 
 (related to the number of species) verify Laplace-like eigenvalue equations, which substantially constraint 
 the asymptotic behaviour and the form of the convex hull polytopes of the Species Scale. 
 It should be interesting to better understand why
  this Laplace conditions, which formally have some resemblance to Heat Equations, are present to begin with.
  We have also seen that the Species Scale may be used as a field-dependent cut-off in EFT's.
  Doing this with the one-loop potential of no-scale moduli in GKP-like scenarios,
  leads to potentials with local minima at the {\it desert points} of the theory and also dS maxima. 
  These minima could be very relevant to the  moduli stabilisation problem.

\vspace*{1.cm}
\textbf{Acknowledgments.}
I would like to thank the organizers of the 
Corfu Summer Institute 2025 "School and Workshops on Elementary Particle Physics and Gravity" for their kind invitation to such a enjoyable conference.
I would like to thank specially my collaborators  Christian Aoufia, Alberto Castellano and Gonzalo F. Casas for a very enjoyable collaboration.
I also thank Anamaria Font,  Alvaro Herraez, Fernando Marchesano, Miguel  Montero and Angel Uranga for useful discussions and input.
This work is supported through the grants CEX2020-001007-S and PID2021-123017NB-I00, funded by MCIN/AEI/10.13039 /501100011033 and by ERDF  'A way of making Europe'.

\vspace*{1.0cm}

%\newpage

\bibliographystyle{JHEP2015}
\bibliography{bibliography}    

%\bibliographystyle{JHEP}
%\bibliography{bib}

%\begin{thebibliography}{99}

%\bibitem{Ferrara:2016ntj}

%\end{thebibliography}

\end{document}